\documentclass[prl,twocolumn,amsmath,amssymb,amsfonts,superscriptaddress,longbibliography]{revtex4-1}

\usepackage{graphicx}
\usepackage{hyperref}
\hypersetup{colorlinks=true}

\usepackage{amsmath}
\DeclareMathAlphabet{\mathpzc}{OT1}{pzc}{m}{it}

\newcommand{\nair}{\textrm{air}}
\newcommand{\npl}{\textrm{pl}}
\newcommand{\nouter}{\textrm{outer}}

\newcommand{\ndecay}{\textrm{decay}}

\newcommand{\nmax}{\textrm{max}}
\newcommand{\nemax}{\textrm{eV,max}}

\newcommand{\be}{\begin{equation}}
\newcommand{\ee}{\end{equation}}

\begin{document}

\title{Excitation of Giant Surface Waves During Laser Wake Field Acceleration}


\author{Travis Garrett}
\affiliation{Air Force Research Laboratory, Directed Energy Directorate, Albuquerque, NM 87123, USA}

\author{C. V. Pieronek}
\affiliation{Air Force Research Laboratory, Directed Energy Directorate, Albuquerque, NM 87123, USA}

\author{E. Rockafellow}
\affiliation{Institute for Research in Electronics and Applied Physics and Department of Physics, University of Maryland, College Park, Maryland 20742, USA}

\author{Oliver Sale}
\affiliation{Leidos Innovations Center, Albuquerque, NM 87106, USA}

\author{Sahir Virani}
\affiliation{Air Force Research Laboratory, Directed Energy Directorate, Albuquerque, NM 87123, USA}

\author{J. E. Shrock}
\affiliation{Institute for Research in Electronics and Applied Physics and Department of Physics, University of Maryland, College Park, Maryland 20742, USA}

\author{B. Miao}
\affiliation{Institute for Research in Electronics and Applied Physics and Department of Physics, University of Maryland, College Park, Maryland 20742, USA}

\author{A. Sloss}
\affiliation{Institute for Research in Electronics and Applied Physics and Department of Physics, University of Maryland, College Park, Maryland 20742, USA}

\author{Jennifer Elle}
\affiliation{Air Force Research Laboratory, Directed Energy Directorate, Albuquerque, NM 87123, USA}

\author{H. M. Milchberg}
\affiliation{Institute for Research in Electronics and Applied Physics and Department of Physics, University of Maryland, College Park, Maryland 20742, USA}


\begin{abstract}

We have detected the presence of very high intensity surface waves  
that are excited during plasma waveguided laser wakefield acceleration.
Wakefield acceleration can be enchanced by the introduction of 
an ``all optical" plasma waveguide that confines 
and guides a laser pulse at the optimal intensity over long distances,  
producing quasimonoenergetic multi-GeV electron bunches.
However strong pulses of radio frequency radiation (RF) are also produced, and   
particle in cell simulations show why: 
a continuous stream of multi-MeV electrons are also ejected radially from the plasma 
due to nonlinear wave breaking, 
and these excite and copropagate coherently with a giant 
cylindrical Sommerfeld surface wave. 
Laboratory measurements, simulations, and analytic approximations  
all converge on a 20 J laser pulse exciting  
a 1 Joule, 400 GW broadband THz surface wave, 
with a peak electric field strength of 35 GV/m.

\end{abstract}

\maketitle

The invention of chirped pulse amplification  
and high intensity femtosecond scale laser pulses 
\cite{strickland1985compression,maine1988generation,kmetec19910,perry1994terawatt} 
have stimulated many  
scientific disciplines and applications 
\cite{hentschel2001attosecond,krausz2009attosecond,
liu1997laser,sekundo2011small,macchi2013ion,di2012extremely,fedotov2023advances},
including laser wakefield acceleration (LWFA) \cite{esarey2009physics}, 
atmospheric filamentation \cite{braun1995self,couairon2007femtosecond}, and the generation of strong 
broadband THz radiation \cite{leitenstorfer20232023,teramoto2018half,shao2025efficiently}. 
We report in this letter on the discovery 
that very high intensity surface waves 
are excited during plasma waveguided LWFA \cite{sprangle1992interaction,durfee1993light}, 
in close analogy to a process that occurs during femtosecond filamentation 
\cite{zhou2011measurement,englesbe2018gas,garrett2021generation,garrett2025detection}. 
Simulations and experiments show that these 
surface waves 
(surface plasmon polaritons (SPP) 
\cite{ritchie1957plasma,pitarke2006theory}
on a plasma substrate)
are an intense source of THz radiation, 
as they convert a significant fraction of the laser energy 
into a broadband THz pulse.  

The key idea behind the original proposal for LWFA \cite{tajima1979laser} 
is that plasmas can support very large accelerating fields ($|E| \sim$ 100 GV/m), 
some $10^3$ times greater than in traditional LINACs. 
In the simplest setup for LWFA, a sufficiently energetic short pulse laser 
is focused into a short gas jet and undergoes relativistic self-focusing in the generated plasma, 
reaching the intensity threshold for ponderomotively expelling electrons 
from the center of the pulse and generating a relativistic plasma wave or wakefield, 
whose axial electrostatic field accelerates electrons loaded 
into the wake via self-injection and wavebreaking.  
The long distance co-propagation of the injected electrons  
with the strong wake fields boosts them to GeV energy scales 
(in contrast to the $\sim$ MeV energies within the laser pulse). 

The addition of a plasma waveguide
\cite{durfee1993light,ehrlich1996guiding,butler2002guiding,
leemans2014multi,miao2020optical,feder2020self,miao2022multi} 
improves the LWFA process by providing a fiberoptic-like channel 
that captures the laser pulse near its beam waist 
and maintains that intensity over many Rayleigh lengths. 
In the first ``all optical" plasma waveguide-based multi-GeV accelerator
\cite{miao2022multi} 
an initial Bessel beam laser pulse was used to rapidly generate 
and heat a long thin column of plasma in an extended supersonic hydrogen gas jet. 
The resulting axially extended cylindrical shock wave expands 
into the surrounding neutral gas, producing a low density plasma core 
surrounded by a higher density radial shell of neutral gas. 
In that experiment and those of this paper, the primary LWFA drive pulse 
(typically with normalized vector potential $a_0 \sim 2$ and pulsewidth $< 100$ fs) 
is injected into the end of this prepared refractive index structure
after a delay of several nanoseconds. 
The radial wings in the very early leading edge of the pulse 
fully ionizes the inside of the shock wall, forming a plasma waveguide  
on the fly, that confines the remainder of the pulse.  
Nitrogen gas can then be added at axial locations along the plasma column 
to enable ionization injection \cite{rowlands2008laser,pak2010injection,mcguffey2010ionization} 
at those positions and subsequent acceleration. 
Recent experiments have confirmed the promise of this technique \cite{shrock2024guided} 
with electron bunches being accelerated to $\sim10$ GeV 
\cite{picksley2024matched,rockafellow2025development}.

A significant side effect has also appeared during these successful waveguided 
LWFA experiments: the generation of unusually intense pulses of RF. 
This radiation has been strong enough to temporarily incapacitate or 
permanently damage a wide range of electronic lab instrumentation including 
computers, motion controllers, oscilloscopes, and vacuum pumps.   
It has been previously noted that intense THz pulses can be produced 
via transition radiation as the electron beams exit the end of the plasma column 
\cite{leemans2003observation,dechard2018terahertz,dechard2019thz}.
However, this new intense RF is still generated during 
when the N2 gas is removed and no GeV electrons are produced. 
In this letter we present new measurements of this intense RF taken during 
LWFA experiments at the ALEPH facility \cite{wang20170}, and show that a novel 
surface wave phenomenon explains their generation.  

Recent experiments and theory 
on RF generation during femtosecond filamentation
\cite{englesbe2018gas,janicek2020length,mitrofanov2021coherently,englesbe2021ultrabroadband,
garrett2021generation,thornton2024boosting,garrett2025detection} 
shed light on the LWFA physics. 
Electrons in a filamentation plasma retain some 
kinetic energy after the laser pulse has passed, typically 1-10 eV 
depending on the laser parameters. The hot tail of the distribution 
expands diffusively from the boundary of the plasma into the surrounding atmosphere, 
thus producing radial current pulse $J_r$ \cite{zhou2011measurement}. 
The amplitude and spectrum of $J_r$ are set by the electron energies,   
an $E_r$ field that grows in response to the current, and electron-neutral collisions
\cite{garrett2021generation}. 

Particle in cell (PIC) simulations \cite{birdsall2004plasma,hockney2021computer} 
reveal that the  $J_r$ current pulse
excites a SPP on the outer boundary of the 
plasma column \cite{garrett2021generation}, 
in particular a transverse magnetic Sommerfeld wave  
\cite{sommerfeld1899ueber,goubau1950surface,stratton2007electromagnetic,
orfanidis2002electromagnetic,pfeiffer1974surface,wang2006dispersion} 
given the cylindrical symmetry 
(see also \cite{op2016single,gong2014electron}). 
The relatively low frequency content of $J_r$  
excites SPPs that are well below the cutoff frequency 
(which is close to the plasma frequency $\omega_{\npl}$, 
$\sim 1$ THz for $n_e \sim 3 \times 10^{22}$ m${}^{-3}$) 
and their group velocity is thus almost $c$. 
The $J_r$ current pulse also translates at nearly $c$ 
behind the laser pulse, 
and it thus co-propagates coherently with the surface wave, 
driving its intensity higher until it saturates. 
At the end of the plasma column most  
of the surface wave detaches \cite{andersen1967radiation,zhang2022towards}, 
becoming forward directed free space RF 
(a time reversal of \cite{stegeman1983excitation}), 
while a fraction of the energy is reflected and travels back up the plasma.

Like filamentation, LWFA also ionizes cylindrical volumes of plasma 
(density $n_e \sim 10^{23}$ m${}^{-3}$)
that grow longer at almost $c$, 
and imparts the electrons with an initial velocity distribution.
However the laser $E$ field strength is roughly 
$10^3$ times stronger during LWFA, 
and the typical electron energies thus $10^6$ higher. 
Most of these MeV scale electrons remain confined within the plasma, 
comprising the Langmuir oscillation, 
but previous PIC simulations
\cite{kalmykov2011electron,galloway2022study}
demonstrate that a subset of them are ejected radially from the plasma.
As these electrons exit the plasma they will drive a far stronger     
$J_r$ current pulse, and excite a much more intense surface wave.    

The Sommerfeld solution has nonzero $E_r$, $E_z$ and $B_{\phi}$ fields,  
which have Bessel function profiles 
inside a cylindrical plasma of radius $r_{\textrm{pl}}$, 
and Hankel function scaling outside.
We measure $E_r$ outside the plasma, 
where the analytic solution is given by:
\begin{equation}
	E_r (r,z,t) = -\frac{\pi r_{\npl} E_0}{2 r_{\nouter}}e^{i(\omega t - hz)}H_1^{(1)}\left(\gamma_{\nair} r\right), 
	\label{eqn_Sommerfeld_Goubau}
\end{equation}
with angular frequency $\omega$, and  
$H_1^{(1)}$ is a Hankel function of the first kind with order 1.
The parameter $\gamma_{\nair}$ is given by  
$\gamma_{\nair}^2 = k_0^2 - h^2$, 
where $k_0=\omega/c$ is the free space wave number,  
$\textrm{Re}(h)$ is the SPP wavenumber and 
$1/\textrm{Im}(h)$ sets an attenuation length scale $L_{\ndecay}$.
We define an outer length scale $r_{\nouter}=1/|\gamma_{\nair}|$: 
for radii $r_{\npl} < r < r_{\nouter}$
the Hankel function is approximately $H^{(1)}_1 \simeq -2 r_{\nouter}/(\pi r)$,
and for $r > r_{\nouter}$ 
it smoothly transitions to an exponential fall off. 
$E_0$ is the peak wave amplitude at the surface of the plasma $r = r_{\npl}$.

\begin{figure}[h!]
  \centering
  \includegraphics[width=0.5\textwidth]{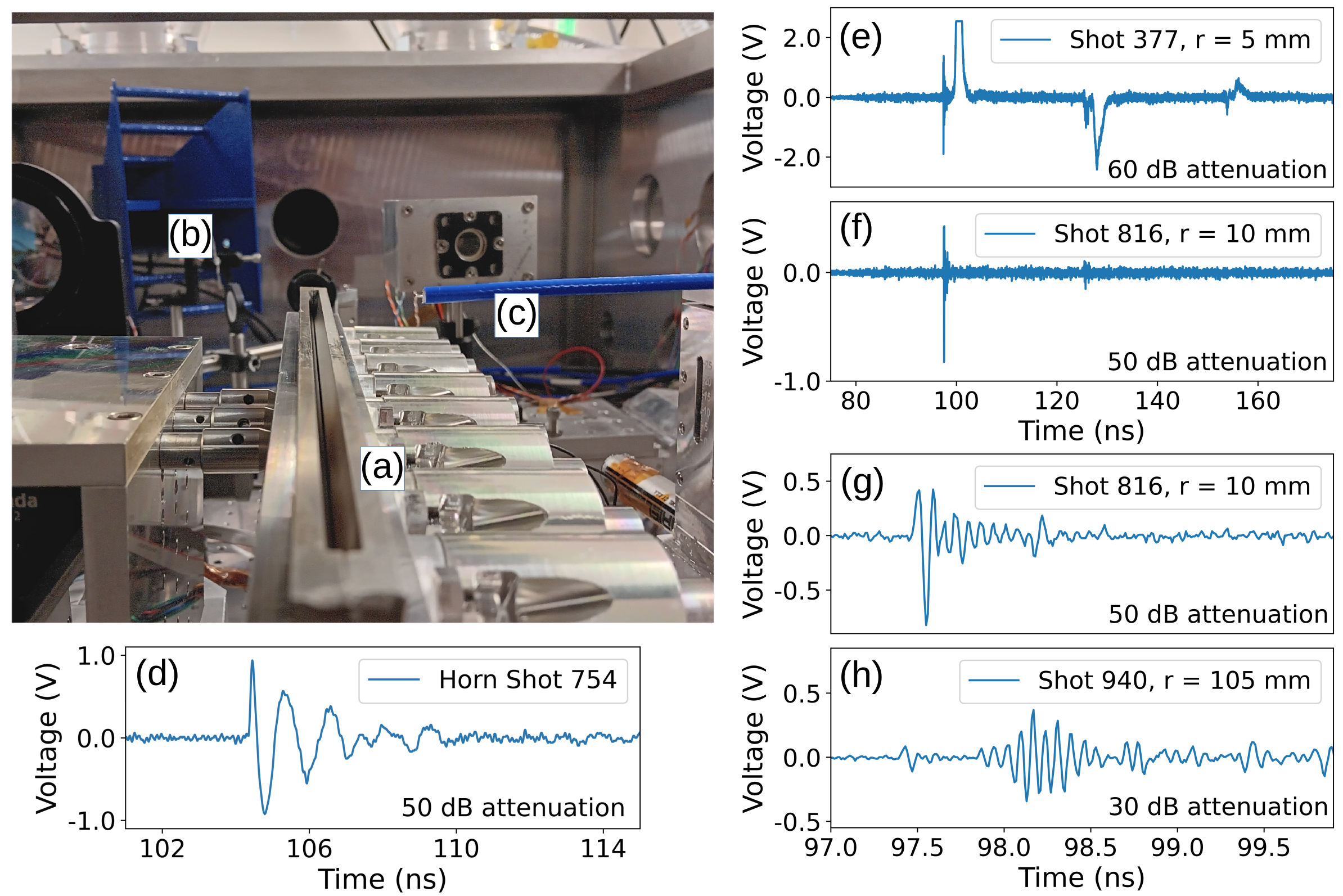}
  \caption{ (a) Down axis view of the gas jet in the LWFA vacuum chamber, 
  at the ALEPH facility. 
  (b) Downstream broadband RF horn. 
  (c) Coax cable converted into a D-dot probe,   
  on a stage that can vary the radius $r$ 
  from 0 to $105$ mm from the plasma that forms above the gas jet.
  (d) Example signal recorded by RF horn, confirming presence 
  of strong RF radiation. 
  (e) Example signal recorded at a distance of $r = 5$ mm from the plasma.  
  In addition to the electromagnetic pulse, 
  a large monopole signal is observed. 
  (f) Example signal recorded at $r = 10$mm during a H2 only run. 
  A strong transient electromagnetic pulse is still generated.
  (g) Zoom in on the transient RF pulse shown in (f). 
  An inverted pulse is seen 0.7 ns after the main pulse, which may
  be due to reflections after the main SPP detaches from the plasma.
  (h) Zoom in on an example signal recorded at the largest 
  separation $r = 105$ mm. 
  At this distance the initial broadband pulse is 
  small compared to the ensuing reflected waves. 
    }
  \label{setup_and_traces}
\end{figure}

The parameter $\gamma_{\nair}$ is found by matching the 
internal Bessel function field profiles to the external Hankel functions. 
Historically a variety of solutions have been developed 
via simplifying assumptions, but the full equations can 
be solved using an iterative method 
\cite{orfanidis2002electromagnetic,garrett2025detection}. 
With an initial guess of
guess of $\gamma_{\nair}^{N=1} = 0.5 k_0$
the subsequent iterations are given by:
\be
\gamma_{\nair}^{N+1} =
\frac{\gamma_{\npl}^N H^{(1)}_1(\gamma_{\nair}^N r_{\npl}) J_0(\gamma_{\npl}^N r_{\npl})}
{\varepsilon_{\npl}(\omega) H^{(1)}_0(\gamma_{\nair}^N r_{\npl}) J_1(\gamma_{\npl}^N r_{\npl})},
\label{iterative_solve}
\ee
with $\gamma_{\npl}^N = \sqrt{(\gamma_{\nair}^N)^2+(\varepsilon_{\npl}(\omega)-1)k^2_0}$,
and plasma permittivity $\varepsilon_{\npl}(\omega) = 1 + \omega_{\npl}^2/(i \omega \nu - \omega^2)$.
The iterative solve typically converges in 10-30 steps \cite{mendoncca2019electromagnetic}, 
thus yielding $r_{\nouter}$ as a function of $r_{\npl}$, $n_e$ and $\omega$. 
Unlike filamentation we assume the electron collision frequency $\nu$
is approximately zero. The electrons at the outer plasma radius 
are initially ionized with low energies, 
but the subsequent arrival of the LWFA SPP will accelerate them 
to $\sim 1/3$ of the speed of light, in which case 
both the electron-ion and electron-neutral 
momentum transfer collision frequencies are negligible. 

\begin{figure}[h!]
  \centering
  \includegraphics[width=0.5\textwidth]{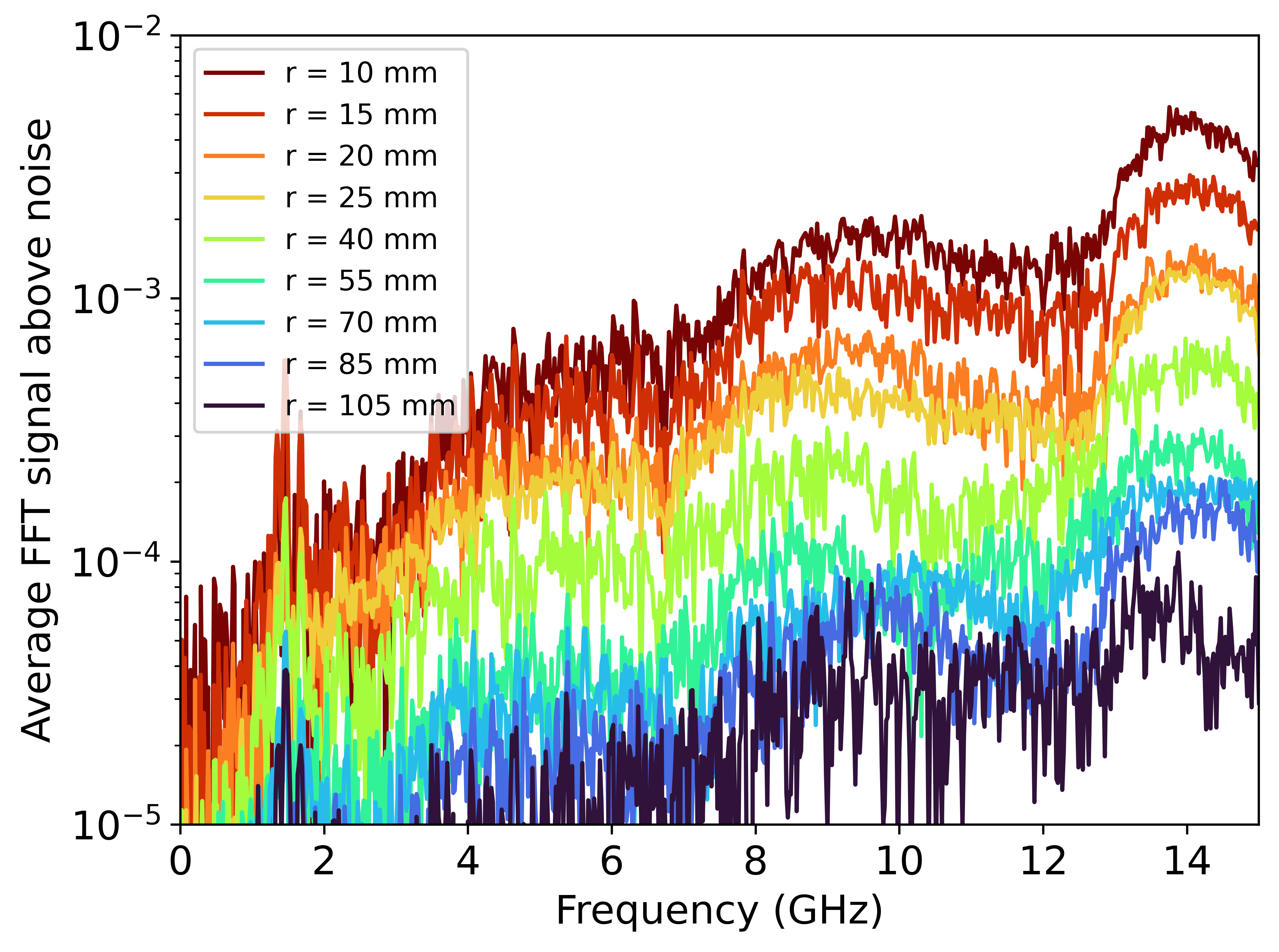}
  \caption{ Average FFTs as a function of D-dot separation 
  from the plasma column, ranging from 10 mm to 105 mm.
  In general there is an approximately linear ramp up in 
  frequency, with additional 
  multi-GHz bumps due to reflections off of the gas jet, 
  and a rapid fall off with increasing separation. 
    }
  \label{fft_plots}
\end{figure}

\begin{figure}[h!]
  \centering
  \includegraphics[width=0.5\textwidth]{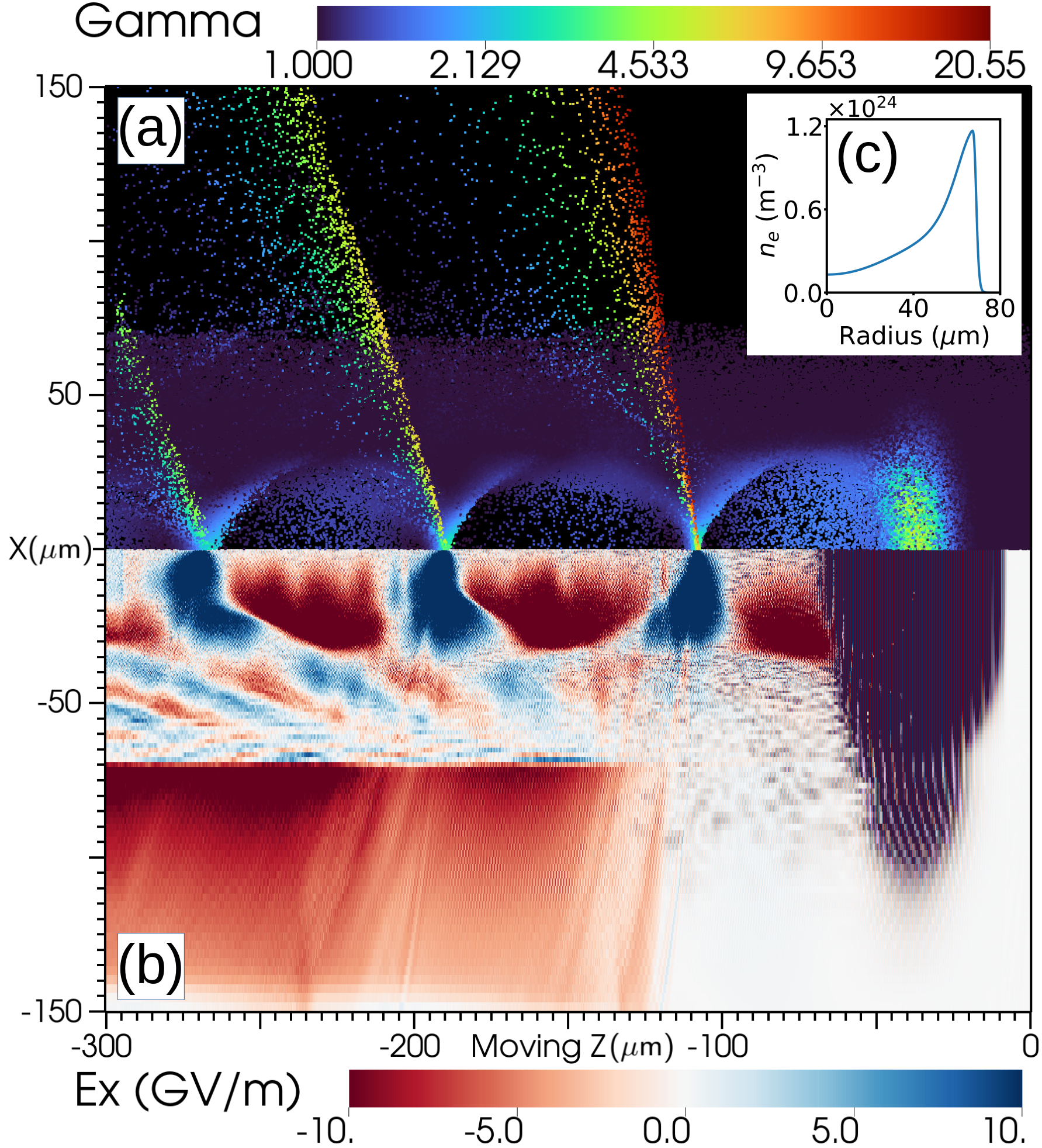}
  \caption{ 
  (a) 2D slice of electron $\gamma$ from a 3D 
  PIC simulation of plasma waveguided $20$ J LWFA, 
  with the $w_0 = $30 $\mu$m beam waist matched to the waveguide. 
  10 MeV electrons are 
  being ejected from the plasma to nonlinear wave breaking.
  (b) $E_x$ field strength from the same simulation, where the color 
  scale has been set to show the strong THz SPP (red) growing 
  on the outer boundary of the plasma.
  After 5 ps of propagation (1.5 mm),
  the peak SPP E field has grown to almost 20 GV/m.
  (c) Radial profile of plasma number density that defines the waveguide. 
   }  
  \label{high_res_pic_sim}
\end{figure}

\begin{figure}[h!]
  \centering
  \includegraphics[width=0.5\textwidth]{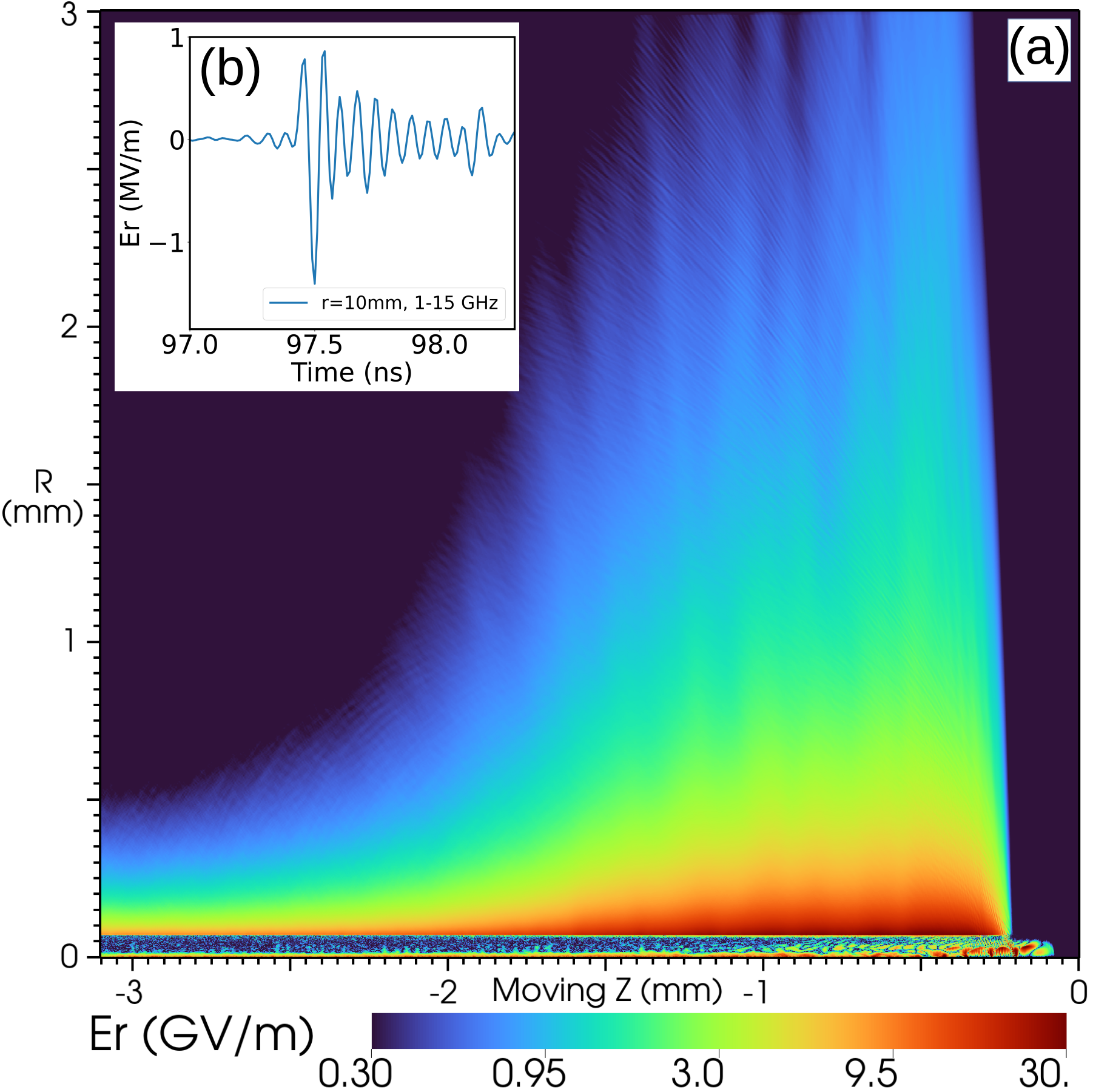}
  \caption{ (a) Log plot of $E_r$ field strength in an axisymmetric PIC 
  simulation which uses a sculpted particle beam to provide the ponderomotive 
  force, leading to the same wave-breaking and surface wave excitation. 
  The 2D $r$-$z$ axisymmetry allows for much larger simulations to run much 
  longer: the pictured wave is after 167 ps (5 cm) of propagation. 
  (b) A measured pulse at $r = 10$ mm in the 1-15 GHz band, corrected for the 50 dB of attenuation,
  transmission loss and converted to electric field strength using the D-dot simulation. 
  Extrapolation to a 1 ps pulse yields a total $E$ field amplitude of $60$ MV/m, in 
  good agreement with simulations which show a $145$ MV/m, 1 ps pulse at $r = 10$ mm. 
    }
  \label{axisym_pics}
\end{figure}

A well matched laser pulse can 
maintain a stable $a_0 \sim 2$  over $10$ cm \cite{shrock2024guided}, providing 
a consistent stream of expelled electrons that 
steadily amplify the Sommerfeld surface wave.
Over time the field strength will saturate (setting the asymptotic $E_0$ value)
as it drives counteracting currents, which occurs roughly 
when the high energy expelled electrons ($K_{\nemax}$) 
are bound by the $1/r$ field. 
For electrons expanding out to $r_{\nmax}$ 
before returning to the plasma, 
the asymptotic surface field strength $E_0$ is approximately: 
\be
E_0 = K_{\nemax} / ( r_{\npl} \ln (r_{\nmax}/r_{\npl}) ).
\label{E0_scaling}
\ee 
For instance, $K_{\nemax}$ = 10 MeV electrons 
that expand out $r_{\nmax} \sim 5$ mm yields $E_0 \sim 35$ GV/m,
in good agreement with the simulations seen 
in Figs \ref{high_res_pic_sim} and \ref{axisym_pics}. 

With lessons from the filamentation radiation in mind we took measurements of  
the RF generated during plasma waveguided LWFA at the ALEPH facility.
Initial measurements inside the vacuum chamber were taken with a 
broadband RF horn (Fig \ref{setup_and_traces}), 
which confirmed that very intense pulses of RF were being generated, 
with the same radial polarization seen during filamentation \cite{garrett2021generation}. 
We then focused on measuring surface waves in the vicinity of the plasma 
column, adapting the technique developed for filamentation \cite{garrett2025detection}. 
A coax cable was converted into a D-dot probe, 
and placed near the midpoint 
of the plasma column on a motorized stage that could scan in the radial direction, 
ranging from 0 mm away from the plasma to 105 mm.
The combination of the internal coax, vacuum feed-through, and external coax 
were characterized with a Rohde and Schwarz Vector Network Analyzer (VNA) 
to determine the frequency dependent transmission loss 
from the D-dot to the oscilloscope. 

\begin{figure}[h!]
  \centering
  \includegraphics[width=0.5\textwidth]{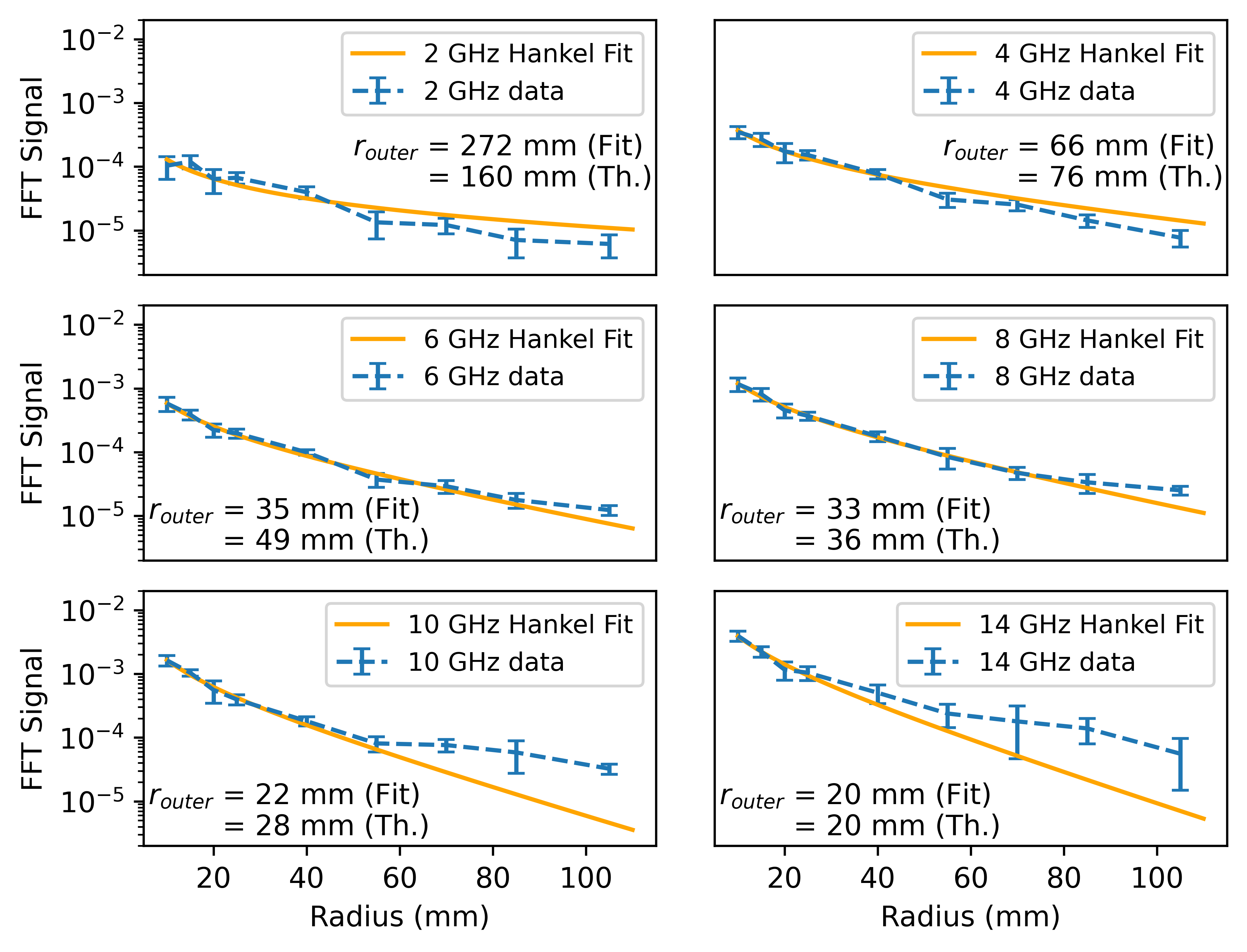}
  \caption{ FFT signal in different frequency bins as 
  a function of radius, and fit to the expected Hankel 
  function profile (which scales as $\sim 1/r$ for $r < r_{\nouter}$
  and $\sim e^{-r}$ for $r > r_{\nouter}$). 
  The data fit values (Fit) for the 
  outer length scale $r_{\nouter}$ are quite close to the 
  theoretical (Th.) predictions as given by (\ref{iterative_solve}) 
  for a plasma column with radius $r_{\npl} = 70$ $\mu$m and 
  electron density $n_e = 2 \times 10^{23}$ m${}^{-3}$.
    }
  \label{hankel_fits}
\end{figure}

As with the horn, very large transient pulses were detected, 
and strong attenuators were added between the coax  
and the oscilloscope, ranging from 60 dB at the smallest 
separations to 30 dB at the largest. 
In addition to the strong electromagnetic signals, at close 
separations ($r = 5$mm) large monopole signals were frequently detected 
(along with subsequent coax reflections): see Fig \ref{setup_and_traces}.
We suspect at this distance that transversely expelled electrons  
are directly colliding with the D-dot probe and depositing charge 
in the metal core, resulting in a monopole pulse as 
the charge returns to ground. 
In addition to this unexpected signal, the noise floor in the 
recorded traces rise in amplitude about 20 ns before the primary signal 
arrives: this may be due to a x-ray background   
that begins arriving at the oscilloscope before 
the RF signal which travels along the curved coax line.

At larger radial separations (Fig \ref{setup_and_traces}) 
the monopole signal disappears, 
leaving only the sharp electromagnetic transient. 
This transient pulse is excited whether Nitrogen is added 
to the gas or not, although the amplitude is 
larger on average when it is included.  
Zooming in we see the primary pulse is broadband, 
and it is consistently followed  
by a smaller inverted secondary pulse about 0.7 ns later, 
which is consistent with some reflected waves returning back upstream 
after the primary SPP detaches from the end of the plasma. 
As we progress out to larger radii the primary 
signal amplitude falls off significantly 
(we reduce the amount of attenuation to preserve resolution), 
and by $r = 105$ mm the subsequent reflected waves 
are significantly louder than the primary.

We next take FFTs of the oscilloscope traces. 
Given the relative magnitude of the later 
reflections at large radii, we truncate the time series data 0.5 ns after 
the primary pulse is detected (the precise form of the truncation 
doesn't significantly change the results). We also use the quiescent 
first half of the truncated time data as a measure of the noise, and subtract 
this from the second half which contains the pulse. Averaging the FFTs together 
for all shots taken at the same radius (and adjusting for varying attenuation)
yields Fig \ref{fft_plots}. In general the spectra ramp up roughly 
linearly with increasing frequency, which is consistent with the RF signal being the 
low frequency tail of a high frequency broadband pulse, and they also fall 
off quickly with increasing radius. Additional multi-GHz bumps can be seen 
on top of the linear trend, which are likely due to reflections 
off of the gas jet assembly below the plasma, while the spike below 
2 GHz stems from a resonance in the D-dot probe. 

The FFTs are then split into 2 GHz bins and averaged, and then 
plotted as a function of radius, see Fig \ref{hankel_fits}. 
In general these are very well fit by the expected Hankel 
function dependence (\ref{eqn_Sommerfeld_Goubau}), 
as was the case during filamentation \cite{garrett2025detection}. 
The fit values for the $r_{\nouter}$ parameter are also very close, 
both in overall magnitude and frequency scaling,  
to the predictions given by (\ref{iterative_solve}) for a plasma with
our plasma parameters. We note that there is extra signal 
at large radii for the higher frequencies, 
which is consistent with free radiation 
being excited by the $J_r$ current pulse in addition to the surface wave \cite{gong2014electron}.

The excellent agreement of the measured radial profiles with the  
Sommerfeld SPP predictions supports the view that energetic electron currents are 
exciting a giant SPP. 
We next perform electromagnetic PIC simulations 
\cite{yee1966numerical,birdsall2004plasma,hockney2021computer,villasenor1992rigorous,boris1970relativistic} 
of the LWFA process, using the ICEPIC code \cite{peterkin2002virtual} 
with moving window reference frame that tracks the laser pulse.
A linearly polarized 20 J, 800 nm, 65 fs FWHM Gaussian laser pulse with a
30 $\mu$m beam waist is  
driven into a 70 $\mu$m radius plasma waveguide (see Fig \ref{high_res_pic_sim}) 
where the radial profile is determined by hydrodynamic 
simulations of the Bessel beam driven cylindrical shock wave 
\cite{miao2024benchmarking}. 
The plasma waveguide does an excellent job of smoothly guiding the pulse 
and holding its intensity and shape nearly constant over 5 ps of evolution, 
driving a blow out bubble in its wake.  
Critically, nonlinear plasma wave breaking 
\cite{dawson1959nonlinear,bulanov1997transverse,esarey2009physics} 
at the end of the first bubble consistently ejects a cone of MeV 
scale electrons out of the plasma column, with a peak 
kinetic energy $K_{\nemax}$ of 10 MeV: 
see Fig \ref{high_res_pic_sim}.
Wave breaking can also been seen expelling lower energy electron jets 
from the subsequent Langmuir oscillations.  
The radial current pulse $J_r$ contained in these conical 
jets excites a massive Sommerfeld SPP
(seen in red in Fig \ref{high_res_pic_sim}, with a 
field strength of $\sim 20$ GV/m at 5 ps), 
which grows steadily larger with propagation distance.

In contrast to other LWFA simulations where the transverse ejection 
of electrons is more sporadic, the stabilization of the laser pulse 
by the plasma waveguide also stabilizes the ensuing plasma 
oscillations, leading to highly regular and energetic wave breaking. 
The peak energy of the ejected electrons grows rapidly with 
increasing $a_0$, and at higher values the ponderomotive force 
directly expels charge as well. At lower $a_0 < 1$ values the clean 
wave breaking at the end of the first oscillation disappears, 
but returns in a turbulent form in later oscillations due to 
increasing wavefront curvature \cite{bulanov1997transverse}. 

The 3D LWFA simulations are computationally expensive,
so we have also performed large scale 2D $\hat{r}$-$\hat{z}$ axisymmetric 
PIC simulations within the same plasma waveguide. 
The laser pulse is replaced with 
a particle beam (traveling at $\gamma = 120$) that has been tuned 
to provide the same ponderomotive impulse,
with a Gaussian radial profile (30 $\mu$m waist)
and a central density of $1.7 \times 10^{23}$ m${}^{-3}$.
The driver particles have the same charge as electrons,
but their mass is increased to $9.1 \times 10^{-25}$ kg 
to maintain a constant ponderomotive force and steady supply 
of $\sim 10$ MeV wave breaking electrons.
An example simulation can be seen in Fig \ref{axisym_pics}: after 
167 ps of evolution (5 cm) the giant surface wave is saturating, 
with 600 MV/m fields extending out to 3 mm.

We finish with 3D conformal ICEPIC calibration simulations
\cite{peterkin2002virtual,merewether1980implementing,taflove2005computational,
zagorodnov2003uniformly,xiao20083,kobidze2010implementation,berenger1994perfectly}
of the D-dot probe being excited by these surface waves,  
which enables the amplitude of the experimental 
traces to be compared with simulated SPPs. 
A 1 ps test wave with an $E$ field amplitude of 1 V/m 
impinging upon the D-dot 
is found to excite a $15$ $\mu$V voltage pulse 
(in the 1-15 GHz band)
that propagates up the coax 
(which is converged at a mesh resolution of 40 $\mu$m).
Adding the measured VNA transmission losses and 50 dB of 
attenuation reduces this to a $12$ nV transient pulse at the 
oscilloscope. 
The signals recorded at $r = 10$ mm have an 
amplitude of $0.7$ V (Fig \ref{setup_and_traces}), 
which yields a field strength of $E \sim 60$ MV/m at the D-dot. 
This is in nice agreement 
large scale axisymmetric simulations, where the 
surface waves have a field 
strength of $E \sim 145$ MV/m at $r = 10$ mm, 
and a duration of $\sim 1$ ps.

Giant surface waves excited by stable wave breaking  
in plasma waveguides provide an intriguing 
alternative use for LWFA as a source of intense THz radiation. 
Extrapolation from measurements, PIC simulations, and 
asymptotic approximations (\ref{E0_scaling}) converge 
on a surface field strength of roughly 35 GV/m, which corresponds 
to 1 J of THz energy, and 5\% conversion efficiency from the 
original 800 nm laser pulse. The very high efficiency 
stems from the combination of the plasma waveguide effectively 
converting the laser pulse energy into Langmuir oscillations, 
which transfer a large fraction of their energy to the expelled 
electrons via nonlinear wave breaking, 
and these co-propagate coherently with a SPP, driving it to very high intensity. 
We are interested in directly detecting this THz radiation 
in future experiments, in modifying its properties by changing the laser parameters
and the profile of the waveguide, 
and investigating out-coupling of the THz for applications.

\begin{acknowledgments}

\textit{Acknowledgments}. The authors thank the CSU Aleph Laser Team 
and F. Sorkin, N.Tripathi, and  M. Gupta (UMd) for technical assistance,
and the Air Force Office of Scientific Research (AFOSR) for support
via Laboratory Task No.s FA9550-24RDCOR002 and FA9550-22RDCOR009. 
The UMd team was supported by the DoE (DE-SC0015516, LaserNetUS DE-SC0019076/FWP\#SCW1668, 
and DE-SC0011375), NSF (PHY2010511), Defense Advanced Research Projects Agency 
(DARPA) under the Muons for Science and Security Program, 
and by the Northrup Grumman Corp. E. R. was supported by an NSF Graduate Research Fellowship (DGE 184034).
This work was supported in part by high-performance computer time
and resources from the DoD High Performance Computing Modernization Program.
The views expressed are those of the author and do not necessarily reflect the official policy 
or position of the Department of the Air Force, the Department of Defense, 
or the U.S. government.
Approved for public release; distribution is unlimited. Public Affairs
release approval No. AFRL-2025-2376.

\end{acknowledgments}

\bibliography{../bib_rf,../surface_waves,../WFA}

\end{document}